\documentclass[ica]{iosart2x}

\usepackage {multicol}
\usepackage{mathtools}

\usepackage{mdframed}
\usepackage[english]{babel}
\usepackage[utf8]{inputenc}
\usepackage{flushend}
\usepackage{rotating,graphicx}
\usepackage{amsmath,amssymb,amsfonts}
\usepackage{bm}
\usepackage{mathtools}
\usepackage{blkarray}
\usepackage{array}
\usepackage[caption=false]{subfig}
\usepackage{capt-of} 
\usepackage[table]{xcolor}
\usepackage{color}
\usepackage{dcolumn}
\usepackage{aeguill} 
\usepackage{stackengine}
\usepackage{setspace}
\usepackage[]{siunitx}
\usepackage{url}
\usepackage{float}
\usepackage{orcidlink}
\usepackage{multirow}
\usepackage{tabularx}
\usepackage{tabulary}
\usepackage{longtable}
\usepackage{threeparttable}
\usepackage{supertabular,booktabs}
\usepackage[noend]{algpseudocode}
\usepackage[ruled,vlined]{algorithm2e}
\usepackage{listings}
\usepackage{comment}
\usepackage{hyperref}
\usepackage{tablefootnote}
\usepackage{natbib}
\usepackage{pifont}
\newcommand{\xmark}{\ding{55}}

\pubyear{2024}
\volume{vol}
\firstpage{1}
\lastpage{21}

\usepackage{lineno}

\usepackage{blindtext}
\usepackage{hyperref}
\usepackage{nameref}

\newcounter{mylabelcounter}

\makeatletter
\newcommand{\labelText}[2]{%
\refstepcounter{mylabelcounter}%
\immediate\write\@auxout{%
 \string\newlabel{#2}{{\unexpanded{#1}}{\thepage}{{\unexpanded{#1}}}{mylabelcounter.\number\value{mylabelcounter}}{}}%
}%
}
\makeatother

\begin{document}
\makeatletter
\let\put@numberlines@box\relax
\makeatother
\begin{frontmatter}

\title{
Identification and Explanation of Disinformation in Wiki Data Streams
}
\runningtitle{
Identification and Explanation of Disinformation in Wiki Data Streams
}

\author[A]{\inits{F.D.P}\fnms{Francisco} \snm{de Arriba-Pérez}\ead[label=e1]{farriba@gti.uvigo.es}},
\author[A]{\inits{S.G.M.}\fnms{Silvia} \snm{García-Méndez}\ead[label=e2]{sgarcia@gti.uvigo.es}\thanks{Corresponding author. \printead{e2}.}},
\author[B]{\inits{F.L.}\fnms{Fátima} \snm{Leal}\ead[label=e3]{fatimal@upt.pt}},
\author[C,D]{\inits{B.M.}\fnms{Benedita} \snm{Malheiro}\ead[label=e4]{mbm@isep.ipp.pt}}
\author[A]{\inits{J.C.B.}\fnms{Juan C.} \snm{Burguillo}\ead[label=e5]{J.C.Burguillo@uvigo.es}},

\runningauthor{F. de Arriba-Pérez et al.}

\address[A]{\orgname{Information Technologies Group, atlanTTic, University of Vigo}, \cny{Spain}\printead[presep={\\}]{e1,e2,e5}}
\address[B]{\orgname{Research on Economics, Management
and Information Technologies, Universidade Portucalense}, \cny{Portugal}\printead[presep={\\}]{e3}}
\address[C]{\orgname{INESC TEC}, \cny{Portugal}\printead[presep={\\}]{e4}}
\address[D]{\orgname{Polytechnic of Porto}, \cny{Portugal}\printead[presep={\\}]{e4}}

\begin{abstract}
Social media platforms, increasingly used as news sources for varied data analytics, have transformed how information is generated and disseminated. However, the unverified nature of this content raises concerns about trustworthiness and accuracy, potentially negatively impacting readers' critical judgment due to disinformation. This work aims to contribute to the automatic data quality validation field, addressing the rapid growth of online content on wiki pages. Our scalable solution includes stream-based data processing with feature engineering, feature analysis and selection, stream-based classification, and real-time explanation of prediction outcomes. The explainability dashboard is designed for the general public, who may need more specialized knowledge to interpret the model's prediction. Experimental results on two datasets attain approximately \SI{90}{\percent} values across all evaluation metrics, demonstrating robust and competitive performance compared to works in the literature. In summary, the system assists editors by reducing their effort and time in detecting disinformation.
\end{abstract}

\begin{keyword}
 \kwd{Disinformation}
 \kwd{eXplainable Artificial Intelligence}
 \kwd{Machine Learning}
 \kwd{Natural Language Processing}
 \kwd{Stream processing}
 \kwd{Wikis}
\end{keyword}

\end{frontmatter}

\section{Introduction}
\label{sec:intro}

Alternative news sources, such as popular social media platforms, have changed how information is generated and disseminated. These sources of unverified content constitute a social concern regarding media trustworthiness and data accuracy, negatively impacting reader critical judgment \cite{Caled2022}. In this regard, recent advances in Natural Language Processing (\textsc{nlp}) like Large Language Models (\textsc{llm}s) have made the generation of human-like web content more affordable, accessible, and faster \cite{Alexiadis2021}. Unfortunately, the latter phenomenon can contribute to disinformation \cite{Marcondes2023}, affecting the quality of data freely available online \cite{Gurrapu2022}.

It is critical to discern misinformation from disinformation. The key difference lies mainly in the intent of the data, \textit{i.e.}, if it is misleading or intended to cause harm (\textit{i.e.}, disinformation) or not (\textit{i.e.}, misinformation) \cite{Broda2024}. Consequently, fake news can be conceptualized as a form of disinformation and are strongly related to emotionally charged and sensationalized data that mimics the style of traditional news stories \cite{Zrnec2022}. The differential factor of the target audience should also be considered \cite{zhou2020survey}.

Therefore, provided the amount and pace of new information generation (\textit{i.e.}, news articles, social media posts, wiki pages), automatic data quality and validation methods have attracted the attention of the research community and the industry \cite{Bordel2019}. Such methods explore Artificial Intelligence (\textsc{ai}) knowledge representation schemes, Machine Learning (\textsc{ml}) models either for supervised or unsupervised learning, and \textsc{nlp} techniques \cite{Guo2022}.

Disinformation approaches explore: (\textit{i}) content, (\textit{ii}) social-context (\textit{i.e.}, user interaction behaviors like commenting, following), or (\textit{iii}) both \cite{GarciaMendez2022SIMPAT}. Finally, deep learning-based solutions exploit neural networks \cite{Brand2023}. Note that this research focuses on traditional \textsc{ml} approaches due to their intrinsic interpretability compared to deep learning algorithms. In this regard, the complexity of automatic disinformation detection systems supported by advanced \textsc{nlp} techniques is strongly related to their opacity (\textit{i.e.}, complex for users to understand the results), leading to lower trustworthiness \cite{Toreini2020}.

Regarding wiki pages or articles, their collaborative nature does not come without a cost. The inability to carry out timely and exhaustive reviews of the content added by the crowd produces misleading or even low-quality articles \cite{Leal2022}. In this regard, groups of volunteers have applied offline methods to assess the quality of Wikipedia\footnote{Available at \url{https://en.wikipedia.org}, November 2024.} articles \cite{Bassani2019}. While Wikipedia has been extensively used for fact-checking, fewer studies examined their content analysis \cite{Chernyavskiy2021,Hsu2021,Trokhymovych2021}. The aforementioned \textsc{nlp} and \textsc{ml} techniques can be exploited for that purpose \cite{GarciaMendez2022,Islam2022}. However, in the Chat\textsc{gpt}\footnote{Available at \url{https://chat.openai.com}, November 2024.} era, existing approaches need further development to meet the highly demanding nature of disinformation, particularly regarding interpretability and explainability \cite{Das2023}. 

Interpretability and explainability refer to different ways of understanding the rationale behind \textsc{ai}-based systems \cite{Macas2023,Mercaldo2024}. Initial advances in eXplainable \textsc{ai} (\textsc{xai}) were based on algorithmic interpretability by exploiting feature relevance and local approximations \cite{nannini2024operationalizing}. However, we may conclude that they are correlated to some extent, so some end users can derive meaning from comprehending how the algorithms operate (\textit{i.e.}, interpretability). Still, most will not and benefit from applying \textit{post hoc} \textsc{xai} techniques. The ultimate objective of these strategies is to provide qualitative meaning by linking input information (\textit{i.e.}, the features) with the outcome (\textit{i.e.}, the prediction) to make the model accountable, reliable, and trustworthy \cite{Lisboa2023}. In short, interpretable \textsc{ai} elaborates on how the systems make the prediction, while \textsc{xai} describes the reasons behind the prediction.

More in detail, interpretability refers to the visualization of relevant model parameters (\textit{e.g.}, feature relevance), which requires technological background, while explainability offers component-independent descriptions of the outcome \cite{Das2023,Salazar2024}. Given that the lack of interpretability and explainability jeopardizes user confidence in the prediction \cite{Kotonya2020,Si2023}, detecting disinformation adaptively in real time is necessary \cite{Kozik2022}. Moreover, shortly, explainability will be mandatory in the European Union\footnote{Available at \url{https://eur-lex.europa.eu/legal-content/EN/TXT/?uri=CELEX:52021PC0206}, November 2024.}.

In fact, the Artificial Intelligence Act (\textsc{aia}) draft in April 2021 represented a major regulatory instance on the field of \textsc{ai}, resulting in the final text approved in December 2023\footnote{Available at \url{https://data.consilium.europa.eu/doc/document/ST-5662-2024-INIT/en/pdf}, November 2024.}. It categorizes \textsc{ai} in different levels of risks with different application limits, transparency requirements, and oversight mechanisms. Particular attention is paid to interpretability regarding the system's accuracy, capabilities, limitations, and use instructions. It also recognizes the right to receive clear explanations with a particular mention to the \textsc{gdpr}. Ultimately, the changes in February 2024 require the disclosure of the use of \textsc{ai} in human interactions \cite{nannini2024operationalizing}. Regarding the future legal requirements in the field of \textsc{ai} in the European Union, even though the act entered into force on 1st August 2024, there exists an adaptation period of two years with some exceptions. Of particular relevance are the governance rules and the obligations for general-purpose \textsc{ai} models, which will be applied in the first year. A more flexible deadline exists for regulated products with embedded \textsc{ai} capabilities. Additionally, provided the complexity of the new regulatory framework, the European Commission proposed the \textsc{ai} pact\footnote{Available at \url{https://digital-strategy.ec.europa.eu/en/policies/ai-pact}, November 2024.} to promote early compliance of future implementations.

The rest of this paper is organized as follows. Section \ref{sec:related_work} overviews the relevant competing disinformation solutions in the state of the art. Section \ref{sec:methodology} introduces the proposed solution, while Section \ref{sec:results} describes the experimental dataset, implementations, and set-up, along with the empirical evaluation results. Finally, Section \ref{sec:conclusions} concludes and highlights the achievements and future work.

\section{Related work}
\label{sec:related_work}

Although disinformation is a recurring problem, recent advances in the field of \textsc{nlp} (\textit{i.e.}, \textsc{llm}s such as Chat\textsc{gpt}) have exacerbated the already high volume of publicly available misleading content, as well as its rate of propagation \cite{Pennycook2021}. Consequently, due to the cost of real-time content patrolling, there is a considerable amount of unreviewed content on the web, which may contain misleading data \cite{Bassani2019}. However, few methods in the literature are dedicated to monitoring the data quality of today's most popular information sources, \textit{i.e.}, wiki pages \cite{Furuta2021}. 

Traditional fact-checking relies on manual verification (\textit{e.g.}, FactCheck\footnote{Available at \url{https://www.factcheck.org}, November 2024.}, PolitiFact.com\footnote{Available at \url{https://www.politifact.com}, November 2024.} or Snopes.com\footnote{Available at \url{https://www.snopes.com}, November 2024.}). Their main drawback is scalability, given human evaluation's operational and time limitations \cite{Sharma2019}. Conversely, \textsc{ml}-based disinformation detectors infer content quality from text features (\textit{e.g.}, text length, lexical aspects such as vocabulary usage), exploit graph representations of the editorial review process (\textit{i.e.}, articles and editors are nodes while the corrections are edges) \cite{Bassani2019}, or employ deep neural network models, sacrificing interpretability in favor of classification accuracy \cite{Gurrapu2022}. The latter prevents their practical use and explains the limited number of proposals in academic research that exploit neural strategies \cite{Hsu2021,Petroni2023}.

Unfortunately, most of the existing disinformation detectors are \textit{black boxes}, lacking the ability to explain the prediction outcome \cite{Fu2022}. Thus, there is a need for advanced solutions that address this concern both effectively and efficiently, \textit{i.e.}, in real-time operation and in a way that is comprehensible not only to experts but also to end users \cite{Leal2021}.

The explainability techniques found in the fact-checking literature are based on salience and logic. Saliency-based methods exploit the attention mechanism to highlight relevant information for end users \cite{Shu2019,Wu2020}. More in detail, they are also called visual \textsc{xai} and represent the first attribution approach applied to convolutional networks, for example, for computer vision. Its name comes after detecting, locating, and segmenting salient data by computing the feature-wise importance score \cite{Borys2023}. Attention-based intelligibility is highly popular with \textit{black box} disinformation detectors. They highlight tokens from the content \cite{Shu2019} or extract relevant $n$-grams using self-attention \cite{Yang2019}. The attention-based descriptions analyze the attention weights of neural models to show which inputs are most relevant for the model's prediction. In this strategy, apart from attention weights computation, another critical task lies in generating context vectors. Logic-based solutions describe the outcome using graphs \cite{Denaux2020} or rule mining \cite{Ahmadi2019}. Specifically, rule mining methods allow modeling data phenomena by exploiting numeric association rules such as the Association Rule Mining (\textsc{arm}) technique, which is a condition statement \cite{Veerappa2021}. While attention-based methods are the most popular technique for explaining profound learning predictions, they require advanced knowledge to maximize the information provided \cite{Shu2019}. Moreover, they may increase the model complexity and limit its adaptability and generalization \cite{Jin2022}. In contrast, the rule mining methods may result in highly transparent descriptions but with low readability properties \cite{Brand2023}. A relevant advantage is that rule-based explanations are more straightforward to comprehend \cite{Leal2021}. However, these systems are limited by the coverage of the knowledge base used for fact-checking \cite{Das2023}. In short, current explainability techniques (\textit{e.g.}, attention scores, Local Interpretable Model-agnostic Explanations - \textsc{lime}, and saliency heatmaps) are unsuited for the general public, who do not have the specialized knowledge to interpret their results \cite{Gurrapu2022}. Finally, interpretable methods (\textit{i.e.}, with non \textit{black box} properties) are considered the most straightforward approach toward direct interpretability and explainability \cite{Kotonya2021}.

Focusing on the works that used wiki pages as fact-checking data sources, we must mention the solution by Sathe \textit{et al.}~\cite{Sathe2020}. They created the WikiFactCheck-English\footnote{Available at \url{http://github.com/WikiFactCheck-English}, November 2024.} dataset composed of almost \num{125000} entries from English Wikipedia articles. The authors validated the usefulness of this corpus with basic \textsc{ml} experiences. Moreover, Fu \textit{et al.}~\cite{Fu2022} proposed \textsc{disco}, an explainable disinformation detection system based on a graph neural network model trained with Wikipedia articles. More recently, Brand \textit{et al.}~\cite{Brand2023} proposed \textsc{e-bart} to detect and explain disinformation, using Wikipedia as a knowledge base.

Few solutions address the analysis of wiki articles in the literature, the works by Bassani \textit{et al.}~\cite{Bassani2019}, Furuta \textit{et al.}~\cite{Furuta2021} and Hsu \textit{et al.}~\cite{Hsu2021} are representative examples. Bassani \textit{et al.}~\cite{Bassani2019} proposed an offline supervised \textsc{ml} approach to address Wikipedia article quality. The authors performed experimental tests with multiple classifiers, obtaining the best results with a Gradient Boosting model. Furuta \textit{et al.}~\cite{Furuta2021} designed an offline fact-checking system based on a Bidirectional Encoder Representations from Transformers (\textsc{bert}) model to identify the parts of articles for the editors to revise. The dataset of sentences usually revised in Wikipedia does not differentiate typographical and grammatical errors from misleading content. Consequently, this research does not address disinformation detection. Conversely, Hsu \textit{et al.}~\cite{Hsu2021} presented the offline Pairwise Contradiction Neural Network (\textsc{pcnn}) model based on the Multi-layer Perceptron (\textsc{mlp}) classifier to identify self-contradictory articles in Wikipedia, relying exclusively on content data. More recently, Z. Chen \textit{et al.}~\cite{Chen2022} proposed EvidenceNet based on the Ro\textsc{bert}a model to retrieve evidence claims from Wikipedia and predict if a statement is supported, refuted by the evidence or cannot be verified. Consequently, the system is exclusively based on content-derived data. Similarly, F. Petroni \textit{et al.}~\cite{Petroni2023} presented \textsc{side}, a neural-network-based solution to detect untrustworthy citations on Wikipedia. Particularly, the verification scores are computed using a fine-tuned \textsc{bert} model. Moreover, S. Shivans \textit{et al.}~\cite{shivansh2023cross} presented XFactVer, which exploits \textsc{bert} for semantic similarity analysis to perform automatic cross-lingual fact-checking in Wikipedia. The authors focused on verifying references as in the work by F. Petroni \textit{et al.}~\cite{Petroni2023}. Ultimately, P. Das \textit{et al.}~\cite{das2024language} presented a computational framework based on regular expressions to assess the quality of Wikipedia content. The authors exploited content-agnostic features (\textit{e.g.}, page length, number of references). Moreover, the experimental labeled data resulted from mapping the quality classes used in Wikipedia. Note that all these solutions operate offline and do not provide explainability.

\subsection{Research contribution}
\label{sec:contribution}

Table~\ref{tab:Comparison} compares the current proposal with the most related research, considering profiling, processing, classification, and outcome explanation. As it can be seen, the prominent approach relies on transformer models (\textit{e.g.}, \textsc{bert}), which justifies the few works that exploit dada beyond content features. It is the case of the works by Bassani \textit{et al.}~\cite{Bassani2019} and P. Das \textit{et al.}~\cite{das2024language}. In light of this comparison, the current proposal is the first to explore content and side features to classify wiki pages in full streaming mode, including the relationship between users and wiki articles. Specifically, the stream operation also applies to the word $n$-gram analysis and hyperparameter selection. The experimental results, obtained with two datasets, analyze the performance of the system in three evaluation scenarios using macro and micro metrics (\textit{i.e.}, accuracy, precision, recall, \textit{F}-measure, and processing time). Finally, textual (\textit{i.e.}, natural language) and visual explainability are provided in the dashboard.

\begin{table*}[!htbp]
\centering
\caption{Comparison with related work from the literature.~\label{tab:Comparison}}
\begin{tabular}{lccccc} 
\toprule
\textbf{Authorship} & \textbf{Profiling} & \textbf{Processing} & \textbf{Model} & \textbf{Explainability}\\
\midrule

\multirow{3}{*}{Bassani \textit{et al.}~\cite{Bassani2019}} & Content & \multirow{3}{*}{Offline} & \multirow{3}{*}{Supervised \textsc{ml}} & \multirow{3}{*}{\xmark}\\
& Review\\
& Network\\

Furuta \textit{et al.}~\cite{Furuta2021} & Content & Offline & \textsc{bert} & \xmark\\

Hsu \textit{et al.}~\cite{Hsu2021} & Content & Offline & \textsc{mlp} & \xmark\\

Z. Chen \textit{et al.}~\cite{Chen2022} & Content & Offline & Ro\textsc{bert}a & \xmark\\

F. Petroni \textit{et al.}~\cite{Petroni2023} & Content & Offline & \textsc{bert} & \xmark\\

S. Shivans \textit{et al.}~\cite{shivansh2023cross} & Content & Offline & \textsc{bert} & \xmark\\

P. Das \textit{et al.}~\cite{das2024language} & Side & Offline & Regular expressions & \xmark\\

\midrule

\multirow{2}{*}{\textbf{Proposal}}	& Content & \multirow{2}{*}{Online} & \multirow{2}{*}{Online} & Textual \\
 & Side & & & Visual\\
 
\bottomrule
\end{tabular}
\end{table*}

In contrast to previous approaches \cite{GarciaMendez2023}, this paper expands on disinformation detection from a multidimensional perspective, \textit{i.e.}, considering performance, efficiency, and fairness. Moreover, it incorporates inputs from expert knowledge, also known as human-in-the-loop \textsc{ai} (\textsc{hai}) \cite{Demartini2020}, and an \textsc{llm} to enhance performance and promote the explainability and fairness of the results. Both inputs are integrated into the explainability control panel: the expert knowledge is used to refine the supervised classifier model via reinforcement learning and the \textsc{llm} descriptions to explain classifier predictions via highly coherent human-like generated texts. Ultimately, \textsc{llm}s and the human-in-the-loop approaches can report relevant advantages to disinformation recognition, especially regarding pattern extraction and explainability \cite{Hanafi2022}. We believe its combination has the potential to share the future detection methods provided the understanding capabilities of \textsc{llm}s. However, these models may incorporate underlying misconceptions due to the absence of a comprehensive understanding of the problem and expert data. The human-in-the-loop approach can do it by leveraging expert knowledge of linguistic patterns, language usage, and deep semantic meaning in disinformation content. This approach also helps to limit the hallucination problem of \textsc{llm}s. Consequently, our system can assist editors by reducing their efforts and time regarding disinformation detection.

In summary, the main contributions of our proposal are:

\begin{itemize}
\item Application of streaming techniques for identifying disinformation in wiki data. Accordingly, our solution is updated in each incoming sample, avoiding the costly batch processing that is the prominent approach in the literature.

\item Generation of a wide range of features to model the evolving nature of disinformation and the crowdsourcing scheme, including the analysis of historical data.

\item Use of \textsc{llm} in combination with \textsc{nlp} techniques such as prompt engineering to generate a natural language description of the system rationale.

\item The possibility of integrating an expert-in-the-loop to verify and correct the system outputs to ensure the proposal's trustworthiness, reliability, and accountability.
\end{itemize}

\section{Methodology}
\label{sec:methodology}

This paper proposes an explainable solution to identify disinformation within reviews through stream-based classification. Our system focuses on wiki posts regardless of their topic (\textit{e.g.}, places, products, services). Note that the heterogeneity of the content is modeled with general and dataset-specific features (see Section \ref{sec:feature_engineering}).

\begin{figure*}[!htbp]
\centering
\includegraphics[scale=0.15]{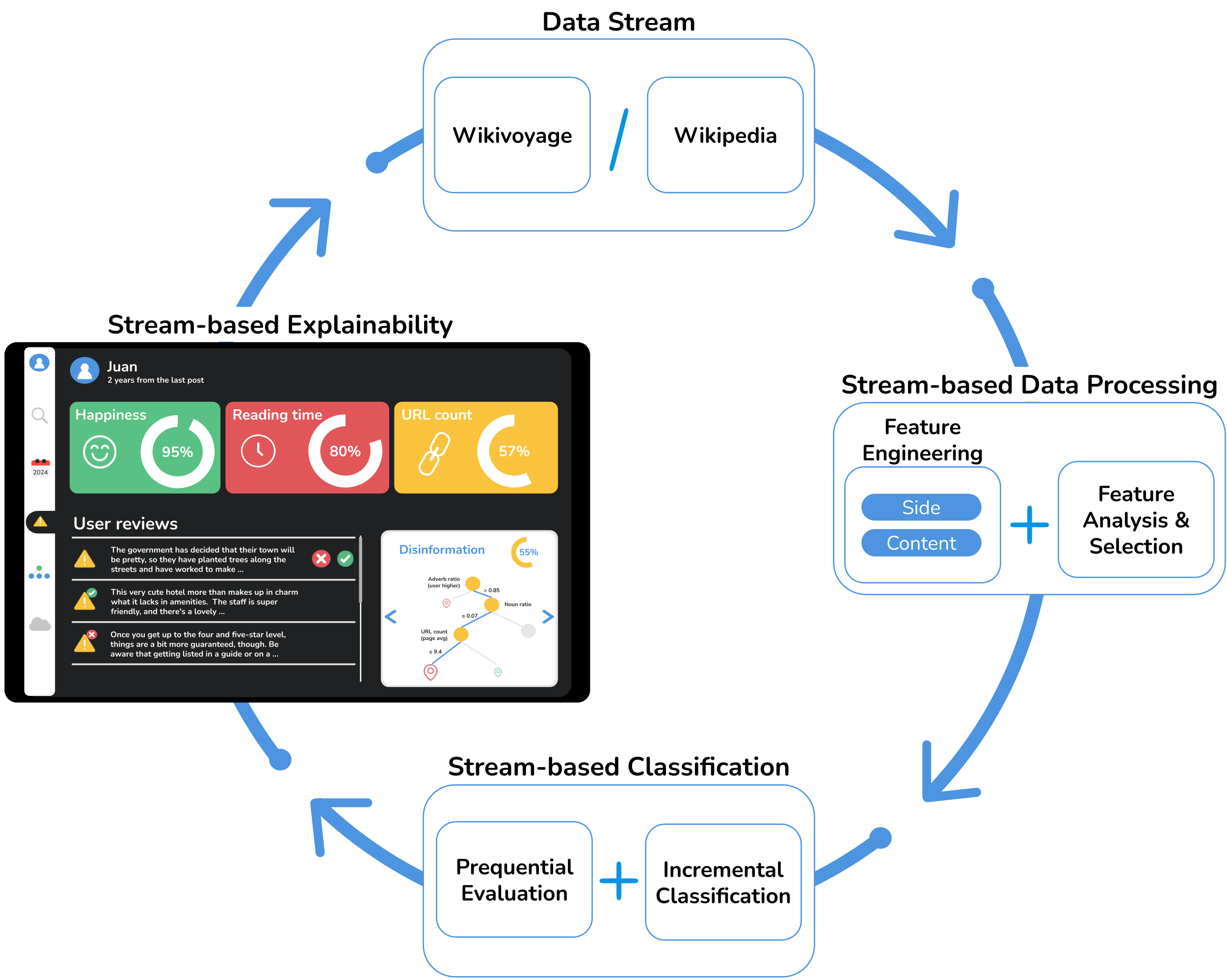}
\caption{\label{fig:scheme}Stream-based classification and explanation of disinformation in wiki streams.}
\end{figure*}

Specifically, we seek to solve a binary classification problem and predict if a post is related to disinformation or non-disinformation content on a streaming basis. Fig. \ref{fig:scheme} shows the proposed scheme composed of several modules that will be described in the subsequent sections: (\textit{i}) stream-based data processing to gather side and content-dependent features, including those based on historical values and their analysis and selection; (\textit{ii}) stream-based classification that evaluates different streaming scenarios (sequential, delayed and batch); and (\textit{iii}) stream-based explainability to provide a natural language description of the model prediction based on an \textsc{llm}.

Note that this section describes the decisions and techniques applied to our approach, while the software, libraries, and configuration parameters are detailed in Section \ref{sec:results}.

\subsection{Stream-based data processing}
\label{sec:data_processing}

Appropriate data processing methods enable filtering non-relevant data for the \textsc{ml} models to perform effectively and efficiently. Data processing in streaming means that the data stream is simulated with individual incoming samples ordered by their timestamp. In this regard, the textual content of the review is preprocessed to remove special characters, numbers, punctuation marks, and subsequent blank spaces. Stop-words are also deleted.

This stage is composed of (\textit{i}) feature engineering to take the most advantage of the experimental data and (\textit{ii}) feature analysis and selection to ensure that incoming data for the latter models is consistent. Note that the contributions of our work, as summarized in Section \ref{sec:contribution}, do not rely on the existing techniques for feature engineering, analysis, and selection but on the way they are integrated into an explainable and online disinformation detection system for wiki data. In this regard, the engineered features result from an in-depth analysis of the crowdsourcing scheme and represent a relevant contribution. Special mention deserved the treatment of historical data.

\subsubsection{Feature engineering}
\label{sec:feature_engineering}

\begin{table*}[!htbp]
\centering
\caption{\label{tab:features}Features explored per experimental dataset.}
\begin{tabular}{cccp{4cm}p{6.5cm}} 
\toprule
\bf Source & \bf Type & \bf ID & \bf Name & \bf Description\\\midrule
\multirow{15}{*}{\rotatebox[origin=c]{90}{Engineering}} & \multirow{9}{*}{\rotatebox[origin=c]{90}{Side}} & 1 & Review count & Number of characters, words, difficult words, and \textsc{url} instances.\\
& & 2 & \textsc{pos} ratio & Ratio of adjectives, adverbs, interjections, nouns, pronouns, punctuation marks, and verbs.\\
& & 3 & Reading time & Content reading time.\\
& & 4 & Flesch score & Readability text score.\\
& & 5 & McAlpine \textsc{eflaw} score & Readability text score for non-native English speakers.\\
\cmidrule{2-5}
& \multirow{6}{*}{\rotatebox[origin=c]{90}{Content}} & 6 & Emotion load & Anger, fear, happiness, sadness, and surprise load.\\
& & 7 & Polarity & Negative, neutral, and positive load.\\
& & 8 & Word2vect & Embedding representation of the content using Word2vect.\\
& & 9 & Word $n$-grams & Single word-grams.\\
\midrule
\multirow{10}{*}{\rotatebox[origin=c]{90}{Wikivoyage}} & \multirow{10}{*}{\rotatebox[origin=c]{90}{Side}} & 10 & Bot flag & It indicates if the author is a bot.\\
& & 11 & Deleted flag & It indicates if the content is hidden.\\
& & 12 & New flag & It indicates that it is the first revision.\\
& & 13 & Revert flag & It indicates if the revision was reverted.\\
& & 14 & Size difference & Difference in the number of characters added and deleted.\\
& & 15 & Article quality & \textsc{ok}, \textsc{wp10b}, \textsc{wp10c}, \textsc{wp10fa}, \textsc{wp10ga}, \textsc{wp10start}, \textsc{wp10stub} probability.\\
& & 16 & Edit quality & False/true damaging \& good faith probability.\\ 
& & 17 & Review quality & \textsc{a}, \textsc{b}, \textsc{c}, \textsc{d}, \textsc{e} probability.\\
\midrule
\multirow{4}{*}{\rotatebox[origin=c]{90}{Wikipedia}} & \multirow{4}{*}{\rotatebox[origin=c]{90}{Side}} 
 & 18 & Article quality & \textsc{ok}, \textsc{wp10b}, \textsc{wp10c}, \textsc{wp10fa}, \textsc{wp10ga}, \textsc{wp10start}, \textsc{wp10stub} probability.\\
& & 19 & Edit quality & False/true damaging \& good faith probability.\\
\\
\bottomrule
\end{tabular}
\end{table*}

Table \ref{tab:features} shows the features engineered for the subsequent supervised learning stage with the \textsc{ml} models. While features 1-9 are generated using \textsc{nlp} techniques, features 10-19 are extracted directly from the wiki. Specifically, the created side features include (\textit{i}) number of characters, words, complex words, and \textsc{url} instances\footnote{Once counted, these elements are removed from the content.}, (\textit{ii}) part-of-speech (\textsc{pos}) ratios (\textit{i.e.}, adjectives, adverbs, interjections, nouns, pronouns, punctuation marks, and verbs), and (\textit{iii}) reading time and readability scores. The new content-related features are built after lemmatization. While lemmatization consists of obtaining a word's original term, considering its meaning (\textit{i.e.}, undoing gender, number), stemming consists of extracting the root term by removing the existing affixes directly. Consequently, the resulting root term may not exist in the language vocabulary and lack contextual meaning. We decided to perform lemmatization instead of stemming, provided its superior performance in the literature regarding natural language mining \cite{mounika2022speech}. These comprise (\textit{i}) emotion and polarity load, (\textit{ii}) word $n$-grams obtained through a dedicated real-time \textit{ad hoc} word $n$-gram extractor, and (\textit{iii}) word embedding representations based on word2vect. More in detail, the \textit{ad hoc} word $n$-gram extractor generates a $n$-size matrix in which every element represents the term and its frequency.

\begin{table*}[!htbp]
\centering
\caption{\label{tab:features_graph}Historical engineered features.}
\begin{tabular}{cp{4cm}p{7cm}} 
\toprule
\bf ID & \bf Name & \bf Description \\\midrule
20 & User post count & Cumulative number of posts per user.\\
21 & User spam tendency & Cumulative ratio of spam posts per user.\\
22 & User posting antiquity & Posting antiquity per user (in weeks).\\
23 & User posting frequency & Weekly posting frequency per user.\\
\multirow{2}{*}{\{24, 61\}} & \multirow{2}{*}{User features} & Incremental average and maximum per user regarding features \numrange{1}{19} in Table \ref{tab:features}.\\
\multirow{2}{*}{\{62, 99\}} & \multirow{2}{*}{Page features} & Incremental average and maximum per page regarding features \numrange{1}{19} in Table \ref{tab:features}.\\
\bottomrule
\end{tabular}
\end{table*}

The posts are made by a user on a specific page based on the content topic. This behavior may be repeated if the wiki identifies a user as a bot or has reverted comments. Likewise, a page that concentrates on numerous reverts is likely arguable. For this reason, the incremental historical values grouped by user and page are calculated for each feature in Table \ref{tab:features}. In this way, our approach is composed of (\textit{i}) raw, (\textit{ii}) engineered, and (\textit{iii}) historical data (see Table \ref{tab:features_graph}).

\subsubsection{Feature analysis \& selection}
\label{sec:feature_analysis_selection}

The feature selection stage becomes challenging in the stream-based scenario since incoming samples may alter the classification variables. The analysis and subsequent selection of the most relevant features (original and engineered) ensure that solely valuable data are submitted to the classification model \cite{Zhang2024}. Features are filtered by a variance thresholding method (\textit{i.e.}, those with low variance are removed). Consequently, this stage aims to reduce the number of features involved in the model training, keeping only those that provide relevant knowledge. Using a dataset partition, the system computes the variance threshold per data stream in the cold start stage, which remains steady when this stage is terminated. The latter directly impacts the system's performance regarding evaluation metrics and time consumed by removing noisy data.

\subsection{Stream-based classification}
\label{sec:classification}

\begin{figure*}[!htbp]
\centering
\includegraphics[scale=0.15]{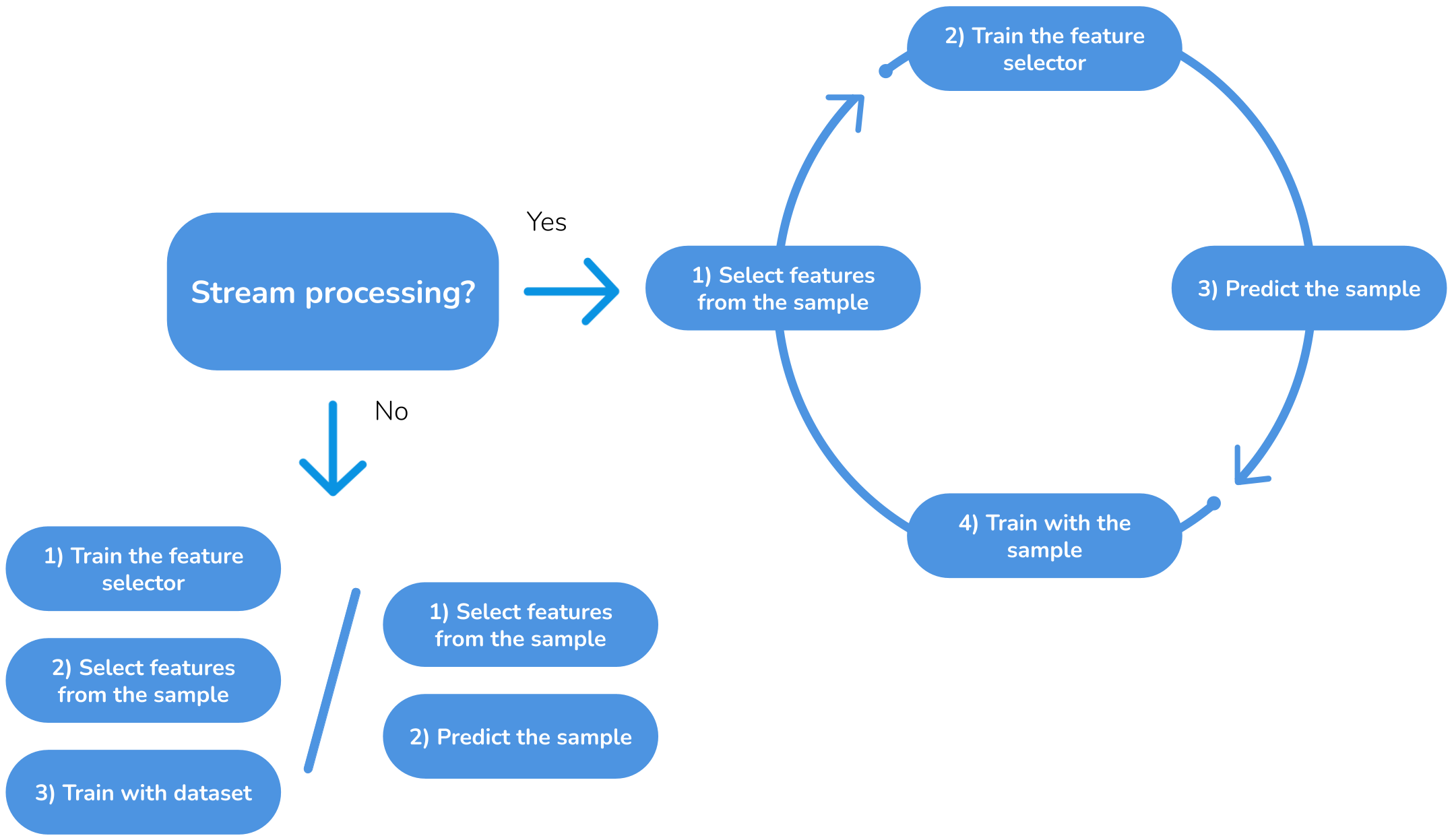}
\caption{\label{fig:streaming_batch} Streaming and batch supervised learning.}
\end{figure*}

Fig. \ref{fig:streaming_batch} shows the pipeline of the supervised learning process performed on a stream basis or batch. In online learning, the first step involves selecting the most relevant features. Moreover, the feature selector is updated with every incoming sample. Then, the resulting category is predicted, and the model is updated with information related to that sample. Conversely, the model can be either trained or used for prediction in batch processing. The training process is slightly different from the online processing. In this case, the selector is trained, and the features are extracted. Ultimately, the training process is performed with the whole dataset, not incrementally as in online processing, preventing a continuous model update. When it comes to batch prediction, the features are first selected.

Our approximation analyzes 3 types of \textsc{ml} classifiers based on Gaussian probabilities, hyperplane, and trees. Note that the latter tree-based models are intrinsically interpretable and were selected based on their promising performance in similar classification problems in the literature and our experimental analyses \cite{Garg2022,patil2022url,de2024exposing,vyas2024real}.

\begin{itemize}
 \item \textbf{Gaussian Naive Bayes} (\textsc{gnb}) \cite{Tieppo2021} implements a stream-based Naive Bayes algorithm based on Gaussian probabilities. It is used for base reference.

 \item \textbf{Approximate Large Margin Algorithm} (\textsc{alma}) \cite{Kang2019} computes the hyperplane maximal margin for linearly separable data in streaming mode.
 
 \item \textbf{Hoeffding Adaptive Tree Classifier} (\textsc{hatc}) \cite{Mrabet2019} consists of a single-tree decision with a branch performance evaluation function for stream-based classification.
 
 \item \textbf{Adaptive Random Forest Classifier} (\textsc{arfc}) \cite{Fatlawi2020} implements an ensemble of \textsc{hatc} with majority voting prediction for stream mining. Specifically, we set a reduced value range for the hyperparameters to prevent a performance reduction regarding the number of estimators or instances of the model (\textit{i.e.}, $models$ hyperparameter), features, and the number of executions per estimator (\textit{i.e.}, using the bagging technique and the $lambda$ hyperparameter). This will ensure the model provides fast responses, \textit{i.e.}, in less than a second. The majority voting is computed taking into account the resulting category provided by each estimator.
\end{itemize}

As the system operates on a streaming basis, the models are updated as samples arrive one by one, ordered by their timestamp (see Fig. \ref{fig:streaming_batch}). First, the features are selected from the specific sample; then, the feature selector is updated to consider the variance of the current features and the previous ones from the previous samples. In contrast to batch learning, the prediction comes first, and then the model is re-trained.

The hyperparameters are optimized at the beginning of the system execution, named the cold start stage, using a configurable set of the initial samples. In this line, the performance of the models is assessed via accuracy, precision, recall, and \textit{F}-measure evaluation metrics, using macro and micro averaging. Note that micro values are especially appropriate in imbalance scenarios and to capture the presence of incorrect predictions (\textit{i.e.}, false positives and false negatives) \cite{Vanacore2023}. These metrics are updated after every new sample. This may result in an imbalance in the target classes at certain moments of the classification process. The latter is the essence of online learning compared to batch learning, which makes the classification problem more challenging. Information about the processing time is also provided.

\subsection{Stream-based explainability}
\label{sec:explainability}

The explainability dashboard provides visual and natural language descriptions of the prediction outcome for algorithmic transparency and increased reliability.

Specifically, our approach generates interpretable predictions using a tree model. These models present bifurcations in each node until reaching the resulting category. Each bifurcation evaluates one of the features, hence allowing the gathering of all involved in the prediction. Then, the description in natural language is generated using an \textsc{llm} exploiting prompt engineering techniques. 

This process is repeated for each of the trees of a Random Forest (\textsc{rf}) classifier, generating as many explanations as the number of trees in the model. Note that our system removes those trees that do not predict the majority category. In addition, our system uses the history of features 1-19 to obtain the quartile in which the user's post is located, along with the three most relevant features detected by the variance selector to display them on the dashboard. Finally, it allows the evaluation of the predictions by an expert, known as the expert-in-the-loop strategy. The samples are tagged and incorporated into the streaming system for model training.

For the natural language descriptions obtained through an \textsc{llm}, we exploited prompt engineering and used the post content as input (see Section \ref{sec:explainability_results} for technical details).

\section{Experimental results}
\label{sec:results}

The experiments were executed on a computer with the following specifications:
\begin{itemize}
 \item \textbf{Operating System}: Ubuntu 18.04.2 LTS 64 bits
 \item \textbf{Processor}: Intel\@Core i9-10900K \SI{2.80}{\giga\hertz}
 \item \textbf{RAM}: \SI{96}{\giga\byte} DDR4 
 \item \textbf{Disk}: \SI{480}{\giga\byte} NVME + \SI{500}{\giga\byte} SSD
\end{itemize}

\begin{figure*}[ht!]
\centering
\includegraphics[width=0.5\textwidth]{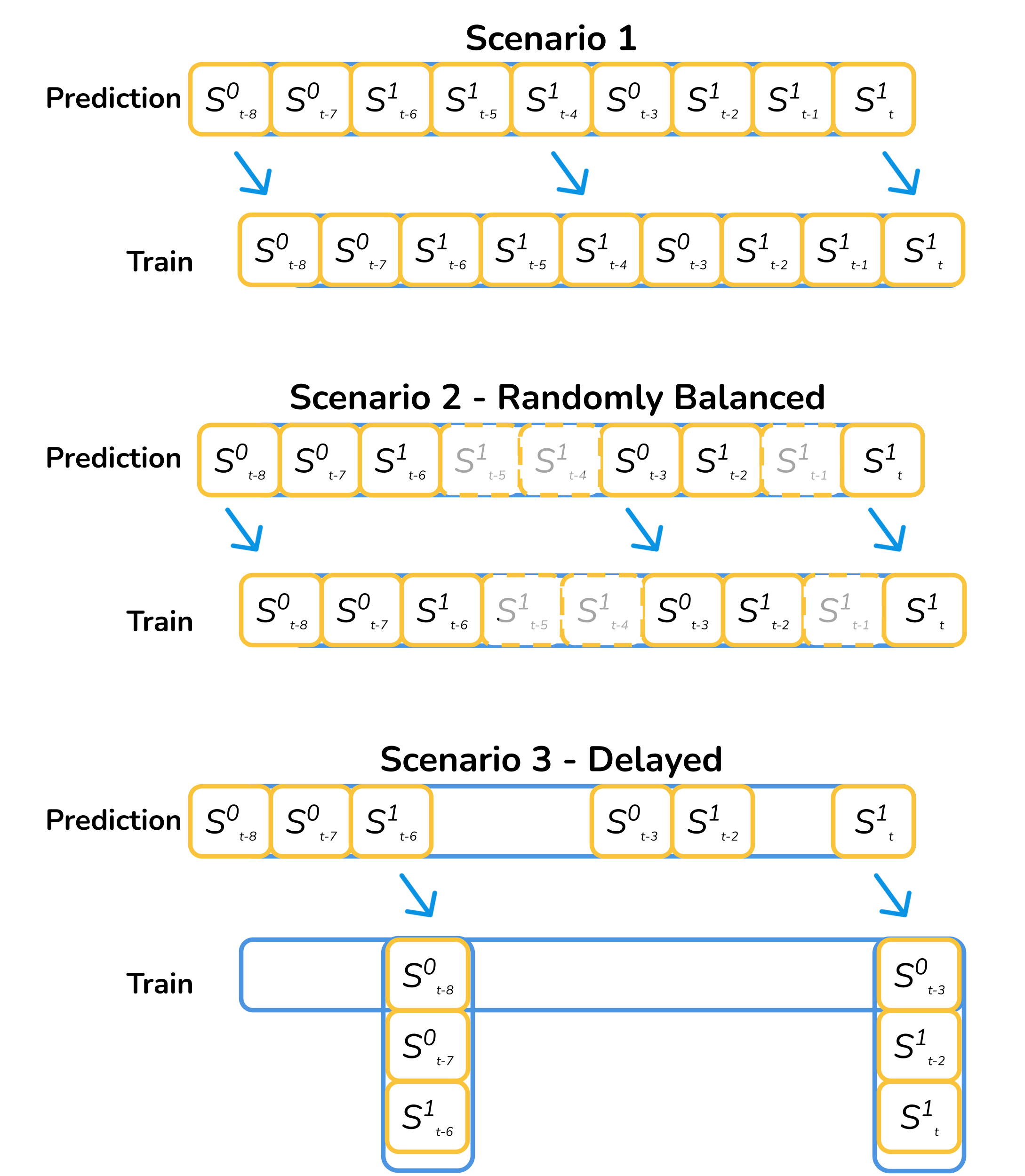}
\caption{\label{fig:scenarios}Data preparation for prediction and training for the three evaluation scenarios considered.}
\end{figure*}

In our stream-based classification problem, the prediction is performed in the order imposed by the timestamp. Considering that the experimental data are unbalanced, different scenarios were designed to assess the system's performance in streaming mode comprehensively.

\begin{description}
 \item \textbf{Scenario 1}. The first $s$ samples are considered for each class, where $s$ is the number of entries for the less frequent class. In this configuration, the dataset is unbalanced by time slots. This means that the samples of a class arrive all together and are not uniformly distributed over time.
 
 \item \textbf{Scenario 2}. The first $s$ samples of the less represented class are considered to respect their distribution in the experimental data. Then, the $s$ samples of the most represented class are randomly distributed to complete $2s$ entries. In this case, classes are balanced so that it is probable to find the same number of samples from both categories in a short period. For this scenario, we use the \texttt{RandomUnderSampler} library\footnote{Available at \url{https://imbalanced-learn.org/stable/references/generated/imblearn.under_sampling.RandomUnderSampler.html}, November 2024.}.
 
 \item \textbf{Scenario 3}. 
 In order to evaluate the system performance in a more realistic layout than predicted with one sample and train with that sample, the training stage is delayed until the $n$-th sample enters the pipeline. Consequently, the data from scenario 2 is used to predict, but the subsequent training stage is performed with $n$ samples in a row. In this scenario, a fixed delay of 100 samples was established. This means the model training takes place when the 100 samples are predicted. This allows us to replicate a real scenario where the expert-in-the-loop labels in fixed-size batches weekly.
\end{description}

Fig. \ref{fig:scenarios} details the data preparation for each scenario. The superscripts indicate the sample target class, and the lowercripts indicate the timestamp.

\subsection{Experimental data}
\label{sec:experimental_data_Set}

Our proposal analyzes a binary classification problem (disinformation versus non-disinformation classes) in wiki data streams. Two experimental datasets were used to analyze the solution's performance comprehensively. The creation of the datasets relies on MediaWiki\footnote{Available at \url{https://www.mediawiki.org/wiki/MediaWiki}, November 2024.} which is an open-source software for wikis initially used on Wikipedia. Therefore, the gathered data was obtained via the MediaWiki Application Programming Interface (\textsc{api}), creating a wiki-based dataset from two wiki-based platforms: Wikivoyage and Wikipedia.

\begin{description}
\item \textbf{Wikivoyage} dataset contains contributions to travel wikis between August 2003 and June 2020. It is composed of \num{285762} entries, distributed between \num{62186} and \num{223576} samples of disinformation and non-disinformation entries, respectively (see Table \ref{tab:distribution}).

\item \textbf{Wikipedia} dataset contains contributions to the Wikipedia platform between September 2006 and December 2023. It is composed of \num{40372} entries, \num{17837} and \num{22535} for disinformation and non-disinformation, respectively (see Table \ref{tab:distribution}).
\end{description}

\begin{table}[!htbp]
\centering
\caption{Disinformation and non-disinformation distribution of entries per experimental dataset.~\label{tab:distribution}}
\begin{tabular}{llS[table-format=6.0]} 
\toprule
\textbf{Dataset} & \textbf{Class} & \textbf{Entries}\\
\midrule

\multirow{3}{*}{Wikivoyage} & Disinformation & 62186\\
& Non-disinformation & 223576\\
\cmidrule{2-3}
& \bf Total & 285762\\
\midrule
\multirow{3}{*}{Wikipedia} & Disinformation & 17837\\
& Non-disinformation & 22535\\
\cmidrule{2-3}
& \bf Total & 40372\\
\bottomrule
\end{tabular}
\end{table}

\subsection{Stream-based data processing}
\label{sec:data_processing_results}

This section reports the techniques used for feature engineering, analysis, and selection and the results obtained. Regarding text processing, special characters, numbers, and punctuation marks\footnote{For numbers and punctuation marks, the removal is performed after engineering side features and before engineering the content ones. Available at \url{https://bit.ly/41dupzh}, November 2024.}, and stop words\footnote{Available at \url{https://gist.github.com/sebleier/554280}, November 2024.} were removed using regular expressions.

\subsubsection{Feature engineering}
\label{sec:feature_engineering_results}

Common classification features encompass side-based (\textit{e.g.}, adjective ratio, reading time) and content-derived data (\textit{i.e.}, embedding representation and single word $n$-grams) as illustrated in Table \ref{tab:features}. The features engineered per dataset are also indicated. Moreover, Table \ref{tab:features_graph} reports the relational features per user and page composed from those listed in Table \ref{tab:features}. Equation \eqref{eq:feature_engineering} represents how the accumulated historical values (for user and page) are calculated, $n$ is the number of sessions and $X[n]$ is the historical feature evaluated in the last $n$ sessions. Expression a) applies to features 20 and 22, while b) to 21, 23-99, and c) to 24-99.

\begin{equation}\label{eq:feature_engineering}
\begin{split}
\forall n \in \{1 ... \infty\}\\ \\
X[n] = \{ x[0],\ldots,x[n]\}. \\
Y[n] = \{y_0[n], y_1[n],\ldots,y_{n-1}[n]\} \mid\\
y_0[n]\leq y_1[n]\leq\ldots\leq y_{n-1}[n], \\
\mbox{where} ; \forall x \in X[n], \; x \in Y[n]. \\ \\
a) sum[n]=\sum_{i=0}^{n} y_i [n]\\
b) avg[n]=\frac{1}{n}\sum_{i=0}^{n} y_i [n]\\
c) \max[n] = \max_{0 \le i \le n} y_i[n]\\
\end{split}
\end{equation}

Particularly, count features (1 in Table \ref{tab:features}) referring to chars and words are computed using the \texttt{len} function in Python. For the latter, the textual content was firstly separated by spaces. For the difficult word count, the \texttt{textstat} \texttt{difficult\_words} function\footnote{Available at \url{https://pypi.org/project/textstat}, November 2024.} was used. Regarding \textsc{url} instances, a regular expression was implemented\footnote{Available at \url{https://bit.ly/3N4GNM3}, November 2024.}. The ratio features (2 in Table \ref{tab:features}) are computed using spaCy\footnote{Available at {\url{https://spacy.io}}, November 2024.} through the \texttt{token.pos\_} function to gather their lexical category. The functions \texttt{reading\_time}, \texttt{flesch\_reading\_ease} and \texttt{mcalpine\_eflaw} of \texttt{textstat} are used to compute reading time and readability scores (features 3, 4, and 5 in Table \ref{tab:features}).

Emotion and polarity load (features 6 and 7 in Table \ref{tab:features}) were calculated using the \texttt{text2emotion}\footnote{Values between 0 and 1. Available at \url{https://pypi.org/project/text2emotion}, November 2024.} and \texttt{TextBlob}\footnote{Values between -1 and 1. Available at \url{https://pypi.org/project/spacytextblob}, November 2024.} functions, respectively. 

Content feature 8 was extracted using the \texttt{word2vect} implementation of the spaCy model \texttt{en\_core\_web\_md}\footnote{Available at {\url{https://spacy.io/models/en}}, November 2024.}. This model transforms each word into a numerical vector of 300 values and thus establishes a proximity relationship. The vector is generated for each word at the sentence level, and its average value is calculated. 

Content feature 9 is generated with an online approximation of \texttt{CountVectorizer}\footnote{Available at {\url{https://scikit-learn.org/stable/modules/generated/sklearn.feature_extraction.text.CountVectorizer.html}}} developed \textit{ad hoc}. Particularly, we use the \texttt{word\_tokenize}\footnote{Available at {\url{https://www.nltk.org/api/nltk.tokenize.html}}, November 2024.}, \texttt{ngrams}\footnote{Available at {\url{https://tedboy.github.io/nlps/generated/generated/nltk.ngrams.html}}, November 2024.} and \texttt{FreqDist}\footnote{Available at {\url{https://tedboy.github.io/nlps/generated/generated/nltk.FreqDist.html}}, November 2024.}. The first extracts the word tokens, the second generates the n-grams object (in our streaming scenario, 1-grams, even though this value can be configurable), and the last allows the frequency computation. A limit is established per sentence for the later single word $n$-grams to control the number of features engineered. To establish the limit above, a cold start of \SI{0.5}{\percent} of the initial samples was used to calculate the second quartile (\textsc{q2}) metric, resulting in four words, which are those with a higher frequency of appearance in the sentence. 

Lemmatization was performed using spaCy with the \texttt{en\_core\_web\_md} model prior to engineering the content features but after engineering the side ones.

In total, \num{89} features were engineered: 80 historical features, 4 content, and 5 side features.

\subsubsection{Feature analysis \& selection}
\label{sec:feature_analysis_selection_results}

The variance threshold was established based on experimental tests as the 90th percentile feature variance value using the \SI{0.5}{\percent} of the initial dataset (cold start) using probabilistic features 15-17 and features 18-19 in the Wikivoyage and Wikipedia experimental data, respectively, resulting in \num{0.067} and \num{0.022}\footnote{These values do not change dynamically.}. The \texttt{VarianceThreshold} method\footnote{Available at \url{https://riverml.xyz/0.11.1/api/feature-selection/VarianceThreshold}, November 2024.} from the \texttt{River} package\footnote{Available at \url{https://riverml.xyz/0.11.1}, November 2024.} enables feature variance computation. This process is dynamic, and the selection of features varies for each sample. In the last sample, 80 and 65 features were discarded for Wikivoyage and Wikipedia datasets, respectively. Reducing features in a streaming system is appropriate and important to avoid resource consumption and keep it at a moderate margin, promoting scalability. Note that the features discarded are those with low variance, so their impact on the model's performance is not expected to be significant. Moreover, removing irrelevant features prevents overfitting in the tree-based classifiers \cite{Oladimeji2024}.

\begin{table}[!htbp]
\centering
\caption{\label{tab:hyperparameter}Hyper-parameter configuration (best values in bold).}
\begin{tabular}{ll} 
\toprule
\textbf{Classifier} & \textbf{Hyperparameter}\\
\midrule
\multirow{3}{*}{\textsc{alma}} 
& alpha = [0.3, 0.5, 0.7, \bf0.9]\\
& B = [0.6, 1.0, 1.4, \bf1.8]\\
& C = [0.6, 1.0, 1.1, 1.4, \bf1.8]\\
\midrule
\multirow{3}{*}{\textsc{hatc}} 
& depth = [None, 50, 100, \bf200]\\
& tiethreshold = [0.9, 0.5, 0.05, \bf0.005]\\
& maxsize = [25, 50, 100, \bf200]\\
\midrule
\multirow{3}{*}{\textsc{arfc}}
& models = [10, 25, 50, \bf75]\\
& features = [sqrt, 25, 50, \bf100]\\
& lambda = [5, 25, 50, \bf100]\\
\bottomrule
\end{tabular}
\end{table}

\subsection{Stream-based classification}
\label{sec:classification_results}

Stream-based {\sc ml} models were incrementally trained and evaluated using an \textit{ad hoc} implementation of \texttt{EvaluatePrequential}\footnote{Available at \url{https://scikit-multiflow.readthedocs.io/en/stable/api/generated/skmultiflow.evaluation.EvaluatePrequential.html}, November 2024.}.

The implementations of the \textsc{ml} models come from the River package\footnote{Due to computational performance, training, and testing were updated every ten slots.}: \textsc{gnb}\footnote{Available at \url{https://riverml.xyz/dev/api/naive-bayes/GaussianNB}, November 2024.}, \textsc{alma}\footnote{Available at \url{https://riverml.xyz/0.11.1/api/linear-model/ALMAClassifier}, November 2024.}, \textsc{hatc}\footnote{Available at \url{https://riverml.xyz/0.11.1/api/tree/HoeffdingAdaptiveTreeClassifier}, November 2024.} and \textsc{arfc}\footnote{Available at \url{https://riverml.xyz/0.11.1/api/ensemble/AdaptiveRandomForestClassifier}, November 2024.}. Table \ref{tab:hyperparameter} shows the hyperparameter optimization ranges used from which the following optimal values were selected for both datasets. Hyperparameter optimization is performed with an exhaustive search over the hyperparameters ranges similar to the operation of the \texttt{GridSearchCV} library\footnote{Available at \url{https://scikit-learn.org/stable/modules/generated/sklearn.model_selection.GridSearchCV.html}, November 2024.} in batch with the \SI{0.5}{\percent} of the original samples as cold start. Moreover, in our binary classification problem, we use 0 and 1 to refer to the non-disinformation and disinformation classes, respectively.

Table \ref{tab:results_wikivoyage} shows the evaluation result for each scenario in the Wikivoyage dataset. The results in scenario 1 are above \SI{81}{\percent}, \SI{78}{\percent}, \SI{65}{\percent}, and \SI{72}{\percent} for the \textsc{gnb}, \textsc{alma}, \textsc{hatc}, and \textsc{arfc} models, respectively. As the problem becomes more challenging and realistic in scenarios 2 and 3, the performance of the models changes significantly. Particularly, for scenario 2, the results are above \SI{66}{\percent}, \SI{52}{\percent}, \SI{74}{\percent}, and \SI{83}{\percent}. A similar performance is obtained in scenario 3. More in detail, the performance of \textsc{gnb} drops (\textit{e.g.}, 0.657 for the recall of the disinformation class in scenario 2). A similar phenomenon is exhibited by the \textsc{alma} model (\textit{e.g.}, 0.481 in the case of the non-disinformation recall in scenario 3). These two models, \textsc{gnb} and \textsc{alma} are the fastest. Moreover, the \textsc{hatc} algorithm attains better results than \textsc{gnb} and \textsc{alma}, and the processing time is in the same magnitude order. However, the values are still variable for this model (\textit{e.g.}, 0.651 to 1.00 between the disinformation class versus the non-disinformation class for the recall metric in scenario 1. Ultimately, the \textsc{arfc} model exhibits a more stable performance even though it is more time-consuming (\textit{i.e.}, values above \SI{80}{\percent} in all evaluation metrics of scenario 2 and 3). Note that this model can process up to 45 samples per second\footnote{\num{62186} non-disinformation and \num{62186} disinformation samples in \SI{2749.25}{\second}.} (0.02 seconds per sample) in scenario 3. This ensures its operation in real time. Moreover, the latter two models, \textsc{hatc} and \textsc{arfc}, are interpretable.

\begin{table*}[htb]
\centering
\caption{\label{tab:results_wikivoyage}Classification results for Wikivoyage dataset (\#0: non-disinformation, \#1: disinformation).}
\begin{tabular}{ccccccccccccS[table-format=4.2]}
\toprule
\bf \multirow{2}{*}{} & \bf \multirow{2}{*}{Model} & \bf \multirow{2}{*}{Acc.} & \multicolumn{3}{c}{\bf Precision} & \multicolumn{3}{c}{\bf Recall} & \multicolumn{3}{c}{\bf \textit{F}-measure} & {\bf Time}\\
\cmidrule(lr){4-6}
\cmidrule(lr){7-9}
\cmidrule(lr){10-12}
& & & Macro & \#0 & \#1 & Macro & \#0 & \#1 & Macro & \#0 & \#1 & {(s)}\\
\midrule
 
\multirow{4}{*}{1} 
& \textsc{gnb} & 82.00 & 0.820 & \bf0.825 & 0.815 & 0.820 & 0.814 & \bf0.826 & 0.820 & 0.819 & 0.821 & 20.00\\
& \textsc{alma} & 77.79 & 0.778 & 0.779 & 0.777 & 0.778 & 0.779 & 0.777 & 0.778 & 0.779 & 0.777 & 10.07\\
& \textsc{hatc} & 82.58 & 0.871 & 0.742 & \bf1.000 & 0.825 & \bf1.000 & 0.651 & 0.820 & 0.852 & 0.788 & 40.88\\
& \textsc{arfc} & \bf85.74 & \bf0.886 & 0.781 & 0.991 & \bf0.857 & 0.994 & 0.721 & \bf0.855 & \bf0.875 & \bf0.834 & 1379.36\\
\midrule

\multirow{4}{*}{2} 
& \textsc{gnb} & 74.32 & 0.750 & 0.711 & 0.789 & 0.742 & 0.828 & 0.657 & 0.741 & 0.765 & 0.717 & 24.26\\
& \textsc{alma} & 52.19 & 0.522 & 0.527 & 0.517 & 0.522 & 0.527 & 0.517 & 0.522 & 0.527 & 0.517 & 12.38\\
& \textsc{hatc} & 77.91 & 0.780 & 0.764 & 0.797 & 0.779 & 0.815 & 0.743 & 0.779 & 0.789 & 0.769 & 64.15\\
& \textsc{arfc} & \bf88.04 & \bf0.884 & \bf0.927 & \bf0.842 & \bf0.881 & \bf0.829 & \bf0.933 & \bf0.880 & \bf0.875 & \bf0.885 & 2701.03\\

\midrule

\multirow{4}{*}{3} 
& \textsc{gnb} & 76.76 & 0.769 & 0.750 & 0.789 & 0.767 & 0.810 & 0.724 & 0.767 & 0.779 & 0.755 & 23.83\\
& \textsc{alma} & 50.64 & 0.507 & 0.513 & 0.501 & 0.507 & 0.481 & 0.532 & 0.506 & 0.496 & 0.516 & 12.33\\
& \textsc{hatc} & 80.87 & 0.809 & 0.824 & 0.794 & 0.809 & 0.791 & 0.827 & 0.809 & 0.807 & 0.811 & 63.65\\
& \textsc{arfc} & \bf87.67 & \bf0.881 & \bf0.925 & \bf0.837 & \bf0.877 & \bf0.822 & \bf0.932 & \bf0.876 & \bf0.871 & \bf0.882 & 2749.25\\

\bottomrule
\end{tabular}
\end{table*}

Table \ref{tab:results_wikipedia} shows the results using the Wikipedia dataset. In this case, the dataset is smaller but quite balanced, allowing us to see each scenario's effect more clearly. Although the behavior is very similar to the previous dataset, it presents a few drops in performance in scenario 3, being \textsc{alma} the most affected classifier. \textsc{arfc} offers stable behavior, with variations of approximately \SI{2}{\percent} concerning scenario 2. Furthermore, \textsc{arfc} has values above \SI{85}{\percent} in all its metrics and scenarios, being especially favorable in scenario 1, with values above \SI{90}{\percent}. It maintains time performance in scenario 3 with the capacity to process 62 samples per second\footnote{\num{17837} non-disinformation and \num{17837} disinformation samples in \SI{574.56}{\second}.} and one sample in 0.02 seconds.

Compared to the results obtained for the Wikivoyage dataset, the ones displayed in Table \ref{tab:results_wikipedia} are more stable regardless of classifiers and scenarios. Based on the discussion provided for Table \ref{tab:results_wikivoyage}, \textsc{arfc} continues to be the best-performing model. Elaborating on scenario 3, the most challenging experiment, the values are close to \SI{80}{\percent}, above \SI{60}{\percent}, above \SI{70}{\percent}, and close to \SI{90}{\percent} for \textsc{gnb}, \textsc{alma}, \textsc{hatc}, and \textsc{arfc} models.

Classifiers based on Gaussian probabilities (\textit{e.g.}, \textsc{gnb}) and hyperplanes (\textit{e.g.}, \textsc{alma}) are more robust to unbalanced data compared to tree-based classifiers (\textit{e.g.}, \textsc{hatc} and \textsc{arfc}) \cite{Alkharabsheh2022}. The reason is that each incoming sample results in the decision path's reconfiguration since it is entirely dependent on the input features. Consequently, \textsc{hatc} and \textsc{arfc} exhibit more significant differences in the recall metric for scenario 1 in the bigger dataset (\textit{i.e.}, Wikivoyage) between the disinformation and non-disinformation classes than  \textsc{gnb} and \textsc{alma}.

\begin{figure}[ht!]
\centering
\includegraphics[width=0.45\textwidth]{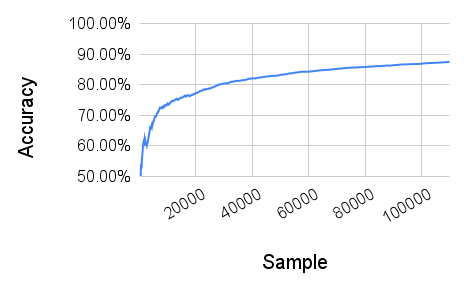}
\caption{\label{fig:acc1}Accuracy evolution in Wikivoyage dataset.}
\end{figure}

\begin{figure}[ht!]
\centering
\includegraphics[width=0.45\textwidth]{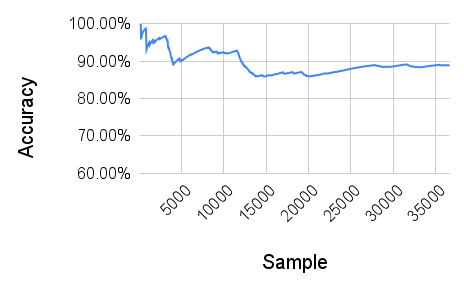}
\caption{\label{fig:acc2}Accuracy evolution in Wikipedia dataset.}
\end{figure}

Fig. \ref{fig:acc1} and Fig. \ref{fig:acc2} display the evolution of accuracy for scenario 3 for Wikivoyage and Wikipedia datasets, respectively. In the first phases, both datasets present their main prediction errors that stabilize and minimize as the sample grows. Due to the delay in training in this scenario, tiny abrupt drops are observed until the model is trained. Especially in the Wikipedia dataset, the curve presents more significant fluctuations due to the smaller number of samples. 

Elaborating on the error analysis for the \textsc{arfc} model in scenario 1, a discussion on the characteristics of the disinformation posts that are more challenging, provided the evolving nature of the system, is appropriate. In this regard and focusing on the Wikivoyage dataset, the incorrectly predicted entries exhibit higher values for feature \textsc{wp10stub}. Similarly, in the Wikipedia dataset, many difficult words and a high value for \textsc{wp10fa} show a detrimental effect. This behavior aligns with the \textsc{ores} article quality probability scheme in which \textsc{stub} refers to concise posts and \textsc{fa} to those who receive a positive score by the reviewers\footnote{Available at \url{https://en.wikipedia.org/wiki/Wikipedia:Content_assessment}, November 2024.}. Consequently, it is reasonable that the system fails when little information is provided, the data are complex (a characteristic of disinformation content \cite{Janicka2019}), or the post scored high in manual verification.

\begin{table*}[htb]
\centering
\caption{\label{tab:results_wikipedia}Classification results for Wikipedia dataset (\#0: non-disinformation, \#1: disinformation).}
\begin{tabular}{ccccccccccccS[table-format=3.2]}
\toprule
\bf \multirow{2}{*}{} & \bf \multirow{2}{*}{Model} & \bf \multirow{2}{*}{Acc.} & \multicolumn{3}{c}{\bf Precision} & \multicolumn{3}{c}{\bf Recall} & \multicolumn{3}{c}{\bf \textit{F}-measure} & {\bf Time}\\
\cmidrule(lr){4-6}
\cmidrule(lr){7-9}
\cmidrule(lr){10-12}
& & & Macro & \#0 & \#1 & Macro & \#0 & \#1 & Macro & \#0 & \#1 & {(s)}\\
\midrule
 
\multirow{4}{*}{1} 
& \textsc{gnb} & 85.53 & 0.855 & 0.854 & 0.857 & 0.855 & 0.855 & 0.856 & 0.855 & 0.855 & 0.856 & 6.28\\
& \textsc{alma} & 81.70 & 0.817 & 0.816 & 0.818 & 0.817 & 0.816 & 0.818 & 0.817 & 0.816 & 0.818 & 2.94\\
& \textsc{hatc} & 83.49 & 0.835 & 0.822 & 0.848 & 0.835 & 0.852 & 0.818 & 0.835 & 0.837 & 0.833 & 13.29\\
& \textsc{arfc} & \bf92.40 & \bf0.924 & \bf0.925 & \bf0.923 & \bf0.924 & \bf0.922 & \bf0.926 & \bf0.924 & \bf0.924 & \bf0.925 & 503.40\\
\midrule

\multirow{4}{*}{2} 
& \textsc{gnb} & 83.15 & 0.833 & 0.857 & 0.809 & 0.832 & 0.800 & 0.864 & 0.831 & 0.827 & 0.836 & 6.13\\
& \textsc{alma} & 72.59 & 0.726 & 0.728 & 0.724 & 0.726 & 0.728 & 0.724 & 0.726 & 0.728 & 0.724 & 3.90\\
& \textsc{hatc} & 81.05 & 0.811 & 0.817 & 0.805 & 0.811 & 0.805 & 0.816 & 0.811 & 0.811 & 0.810 & 12.89\\
& \textsc{arfc} & \bf90.33 & \bf0.904 & \bf0.923 & \bf0.885 & \bf0.904 & \bf0.882 & \bf0.925 & \bf0.903 & \bf0.902 & \bf0.905 & 615.00\\

\midrule

\multirow{4}{*}{3} 
& \textsc{gnb} & 80.82 & 0.809 & 0.831 & 0.788 & 0.809 & 0.778 & 0.839 & 0.808 & 0.804 & 0.813 & 6.11\\
& \textsc{alma} & 63.34 & 0.634 & 0.642 & 0.626 & 0.634 & 0.618 & 0.649 & 0.633 & 0.630 & 0.637 & 2.99\\
& \textsc{hatc} & 72.81 & 0.728 & 0.735 & 0.722 & 0.728 & 0.721 & 0.735 & 0.728 & 0.728 & 0.728 & 13.21\\
& \textsc{arfc} & \bf88.90 & \bf0.889 & \bf0.902 & \bf0.877 & \bf0.889 & \bf0.875 & \bf0.903 & \bf0.889 & \bf0.888 & \bf0.890 & 574.56\\

\bottomrule
\end{tabular}
\end{table*}

\subsection{Stream-based explainability}
\label{sec:explainability_results}

\begin{figure*}[ht!]
\centering
\includegraphics[width=0.6\textwidth]{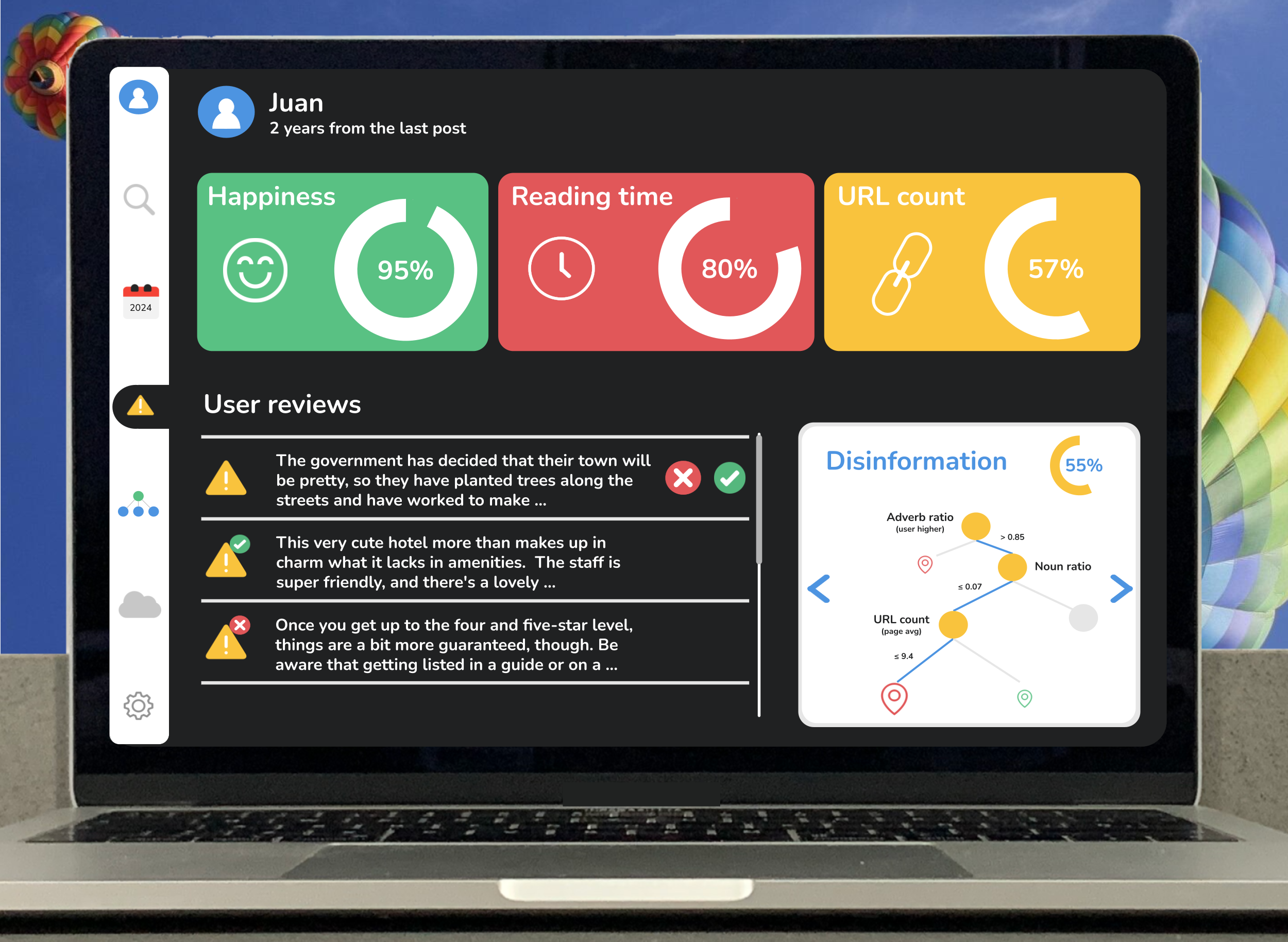}
\caption{\label{fig:dasboard}Explainability dashboard.}
\end{figure*}

\begin{figure*}[ht!]
\centering
\includegraphics[width=0.6\textwidth]{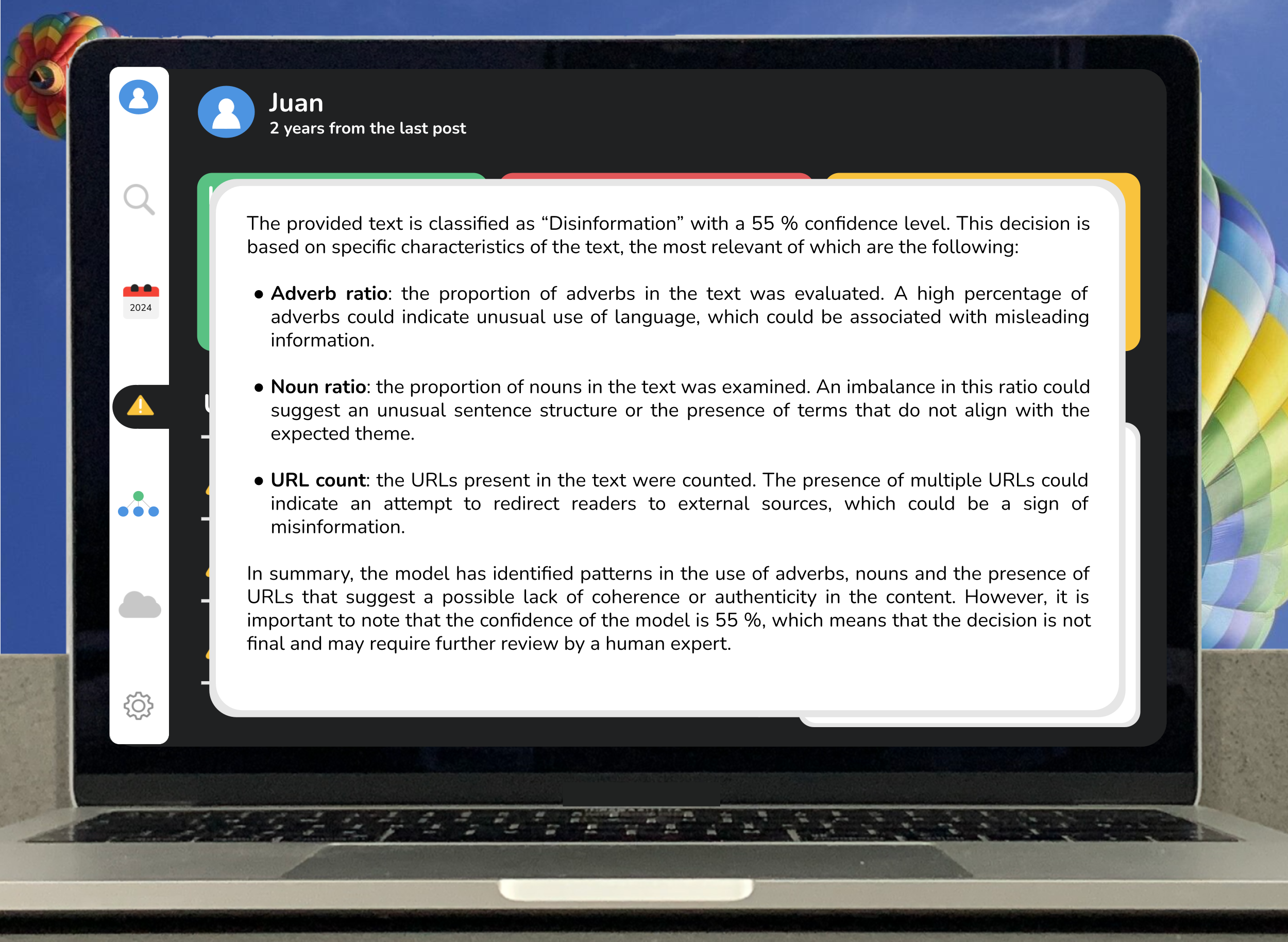}
\caption{\label{fig:explainability}Pop-up explainability of the decision tree path.}
\end{figure*}

The explainability dashboard is presented in Fig. \ref{fig:dasboard}. At the top, information related to the username and the last contribution are displayed. The three most relevant features used for prediction are shown below. The green, yellow, and red colors indicate whether the value of the feature has exceeded quartiles 1, 2, and 3, respectively, concerning all users. The user's contributions are shown at the bottom left. The human-in-the-loop can correct the system by validating or rejecting the predictions in this section. The evaluation option is only available for those not previously corrected or validated. In the bottom right part, the route of a decision tree and the selected features are shown until reaching the predicted category. Note that this decision path visualization applies only to tree-based models (\textsc{hatc} or \textsc{arfc}) in our experiments. The end user can select the trees that result in the majority label, and the natural language description is updated accordingly (see Fig. \ref{fig:explainability}). The prediction confidence percentage based on \texttt{Predict\_Proba\_One} function\footnote{Available at \url{https://riverml.xyz/0.11.1/api/base/Classifier}, November 2024.} is also provided. 

Particularly, for the natural language explanation in Fig. \ref{fig:explainability}, we exploited the \textsc{gpt}3.5-turbo\footnote{Available at {\scriptsize \url{https://openai.com/product}}, November 2024.} model using the default parameters, the following prompt, and the post content. An example of the output provided by the \textsc{llm} is shown in Fig. \ref{fig:explainability}.

\begin{itemize}
 \item Prompt: \textit{Our Machine Learning model has predicted that this text <Text> is classified as <Category> with a confidence of <Confidence>\%. The most relevant path features are: [List of relevant features].
 Generate a human-explainable text that summarizes the decision made by the classifier.} 
\end{itemize}

\begin{table*}[!htbp]
\centering
\caption{\label{tab:comparisonclassification_results} Comparison with the best approach from the literature.}
\begin{tabular}{ccccccccS[table-format=3.2]}
\toprule
\bf {Proposal} & \bf{Technique} & \bf {Accuracy} & \bf {\bf Time}\\
\midrule
Furuta \textit{et al.}~\cite{Furuta2021} & \textsc{bert} & 69.10 & \textsc{na}\\
Hsu \textit{et al.}~\cite{Hsu2021} & \textsc{mlp} & 63.12 & \textsc{na}\\
Z. Chen \textit{et al.}~\cite{Chen2022} & Ro\textsc{bert}a & 76.95 & \textsc{na}\\
F. Petroni \textit{et al.}~\cite{Petroni2023} & \textsc{bert} & 85.69 & \textsc{na}\\
S. Shivans \textit{et al.}~\cite{shivansh2023cross} & \textsc{bert} & 70.52 & \textsc{na}\\
P. Das \textit{et al.}~\cite{das2024language} & Regular expressions & 83.90 & \textsc{na}\\
\bottomrule
\end{tabular}
\end{table*}

\subsection{Discussion}
\label{sec:discussion}

The proposed method addresses the problem of identifying and explaining disinformation in wiki platforms using wiki events as streams. However, since the surveyed works use an offline processing approach, comparing the results may not be straightforward (see Table \ref{tab:Comparison}). 

The Wikipedia article quality assessment solution (\textit{i.e.}, to distinguish those with high or low-quality content) by Bassani \textit{et al.}~\cite{Bassani2019} attained similar results as our proposal (\SI{91.40}{\percent} accuracy with 0.085 Mean Squared Error - \textsc{mse}) with the Gradient Boosting model (see Table \ref{tab:comparisonclassification_results}). Moreover, Furuta \textit{et al.}~\cite{Furuta2021} proposed a fact-checking assistant for Wikipedia. However, the accuracy obtained with the \textsc{bert} model is low (\SI{69.10}{\percent}, \num{23.3} percent points lower than the \textsc{arfc} model in scenario 1 for the Wikipedia dataset). Hsu \textit{et al.}~\cite{Hsu2021} created WikiContradiction to detect self-contraction articles for Wikipedia. The best result obtained with a balanced setting and test-train ratio of \SI{80}{\percent} with the \textsc{mlp} model is \SI{63}{\percent} for accuracy (\num{27} percent points lower than the \textsc{arfc} in scenario 2 for the Wikipedia dataset). Regarding the more recent works, the solution by Z. Chen \textit{et al.}~\cite{Chen2022} attained the best results with the Ro\textsc{bert}a model, however below the \SI{80}{\percent} threshold. Similarly, F. Petroni \textit{et al.}~\cite{Petroni2023} and S. Shivans \textit{et al.}~\cite{shivansh2023cross} proposed to exploit the \textsc{bert} model which provided up to \SI{85}{\percent} accuracy (7 percent points lower than the \textsc{arfc} model in scenario 1 for the Wikipedia dataset). Ultimately, the solution by P. Das \textit{et al.}~\cite{das2024language} based on regular expressions attained similar results to those by F. Petroni \textit{et al.}~\cite{Petroni2023}.

Unfortunately, these works discussed do not provide training time information, operate offline, and lack explainability capabilities.

\section{Conclusions}
\label{sec:conclusions}

Unverified sources of information (\textit{e.g.}, social media platforms, and wiki pages) pose a significant threat regarding trustworthiness and data accuracy, mainly impacting the quality of the information generated and disseminated among end users. Apart from low-quality content, the spread of misleading data and misinformation is particularly relevant. The current data quality and validation methods in the literature cannot carry out timely and exhaustive reviews of the content added by the crowd. Accordingly, this work focused on wiki pages, a popular data source for fact-checking analysis, even though fewer studies examined its content. 

Our solution comprises stream-based data processing with feature engineering (content and side features) and feature analysis and selection, stream-based classification, and real-time explanation of the prediction outcome. The explainability dashboard was intended for the general public without requiring specialized knowledge to interpret the decision path of the classification model. Moreover, the system's performance was evaluated with two datasets in three scenarios, attaining evaluation metrics around \SI{90}{\percent}. Thus enhancing the performance of existing competing solutions in the literature. Summing up, it contributes to reducing the cost of real-time content patrolling of disinformation data in an automatic and scalable way compared to traditional approaches, which rely exclusively on manual supervision.

Future work will leverage the system's modular design to integrate claim verification through evidence retrieval. Some technical design decisions will also be re-evaluated, such as the dynamic adjustment of the variance threshold for feature analysis and selection to address potential changes in feature importance over time (\textit{i.e.}, concept drift), sensitivity analysis on different delays for scenario 3 and alternative techniques such as mutual information and forward selection. Multilingual processing and using an \textsc{llm} for feature engineering will also be explored. The system will be made available to the research community via \textsc{api} to analyze user feedback, for example, regarding the explainability dashboard. Additionally, we plan to perform new experiments using the self-supervised learning strategy \cite{Rafiei2022}, ensemble learning \cite{alam2020dynamic}, and neural algorithms \cite{Rafiei2017} as well as auto-\textsc{ml} approaches.

\section*{Acknowledgements}

This work was partially supported by: (\textit{i}) Xunta de Galicia grants ED481B-2022-093 and ED481D 2024/014, Spain; and (\textit{ii}) Portuguese National Funds through the FCT – Fundação para a Ciência e a Tecnologia (Portuguese Foundation for Science and Technology) as part of project UIDB/50014/2020.

\bibliographystyle{ios1}
\bibliography{bibliography} 

\end{document}